\documentclass[times,letter,conference]{IEEEtran}
\usepackage{cite}
\usepackage[utf8]{inputenc}   
\usepackage[T1]{fontenc}     
\usepackage{geometry}      
\usepackage{epsfig}
\usepackage{graphicx}
\usepackage{enumerate}
\usepackage{listings}
\usepackage{xcolor}

\begin{document}
\author{\IEEEauthorblockN{Daniela Genius}
\IEEEauthorblockA{Sorbonne Universit\'e, LIP6, CNRS UMR 7606, daniela.genius@lip6.fr}
}

\title{Scalability of TTool's AMS extensions: a case study}
\maketitle
\begin{abstract}
Embedded cyber-physical systems (CPS) are commonly built upon heterogeneous digital and analog integrated circuits, including sensors and actuators. 
Less common is their deployment on parallel, NoC based
designs based on general purpose processor cores
of a Multi-processor System-on-chip (MPSoC). Application code has to be run
on the MPSoC for the digital part, and interact with the analog sensors.
We recently proposed a major extension to the design and exploration
tool named TTool, 
 now allowing the design of CPS
 on a high level of abstraction and the generation of
cycle-bit accurate simulations.
We explore the scalability of our 
approach with an automotive case study.
\end{abstract}
\section{Introduction}
Many applications e.g. from robotics, automotive and autonomous systems
require heterogeneous modeling  - including modeling of analog/mixed 
signal (AMS) and radio frequency (RF) features.
In very early design phases, rapid but less precise exploration of the design space is required.
Co-simulation of heterogeneous embedded systems then requires a high-level 
representation that includes high-level models of AMS and RF components. 
In the approach we presented in \cite{RAPIDO2019} software
is run on the digital multiprocessor
system on chip (MPSoC) platform, but has to communicate with the analog part.
Whereas that work took as a starting point an analog application
where only the controller was defined as a task running on a SoC, \cite{MODELSWARD2019} presented a full methodology on a toy example.  
The present paper explores a far more complex case study, 
where a lot more functionality resides in the software tasks. 

The next section presents the ways in which
the design problems problems
are addressed in similar research work.
Section \ref{context} presents the bases.
Some of the questions raised when modeling
larger applications are listed in Section \ref{contribution};
Section \ref{casestudy} illustrates them by a larger case study,
Section \ref{conclusion} concludes the paper.

\section{Related Work}\label{relatedwork}
Well established tools in analog/mixed signal design,
like \textit{Ptolemy II}, \cite{LEE.10}
\cite{PTOL.14}, based upon a data-flow model,
address heterogeneous systems by defining several sub domains 
using hierarchical models. Instantiation of elements controlling
the time synchronization between domains is left to the responsibility
of designers.

Metropolis \cite{METROPOLIS} is based on a high level model
and facilitates the separation of computation from communication concerns.
Metro II \cite{davare2007next} introduces hierarchy and allows \textit{Adaptors} for data synchronization
as a bridge between the semantics of components belonging to different MoCs;
the model designer still has to implement time synchronization.
As a common simulation kernel handles all process execution, MoCs are not well separated. 

Discrete Event System Specification (DEVS \cite{DEVS}), a modular and hierarchical formalism for modeling and analyzing general systems that can be discrete
event or continuous state systems. 

SystemC \cite{SYSTEMC} is a C++ class library which makes it possible to model
(digital) hardware on multiple levels of abstraction.
Among the frameworks based on SystemC are HetSC \cite{herrera2007framework},
HetMoC \cite{zhu2010hetmoc} and ForSyDe \cite{niaki2012formal}, all having the disadvantage that instantiation 
of elements and
controlling the synchronization have to be managed by the designer.

SystemC-AMS extensions \cite{SCAMSUG} is a standard 
describing an extension of SystemC with AMS and RF features \cite{VachouxGE03}.
The usual approach for modeling the digital part of a heterogeneous system with SystemC is to rely on the
 \textit{Discrete Event} (DE) part of SystemC AMS extensions. The \textit{Timed data Flow} (TDF) part adds support for signals where data values are sampled with a constant time step. 

In the scope of the project BeyondDreams \cite{beyond_dreams},
a mixed analog-digital systems proof-of-concept simulator has been developed, 
based on the SystemC AMS extension standard. 
Another simulator is proposed in the H-Inception project \cite{hinception}.  
All of these approaches rely on SystemC AMS code i.e. they don't provide a high-level interface for specifying the application.
Integration with software code for general-purpose CPUs and with an operating system
is however not yet addressed in these approaches.

\section{Context}\label{context}

\subsection{TTool}
\begin{figure*}[hbt]
\centering
\includegraphics[width=0.9\linewidth]{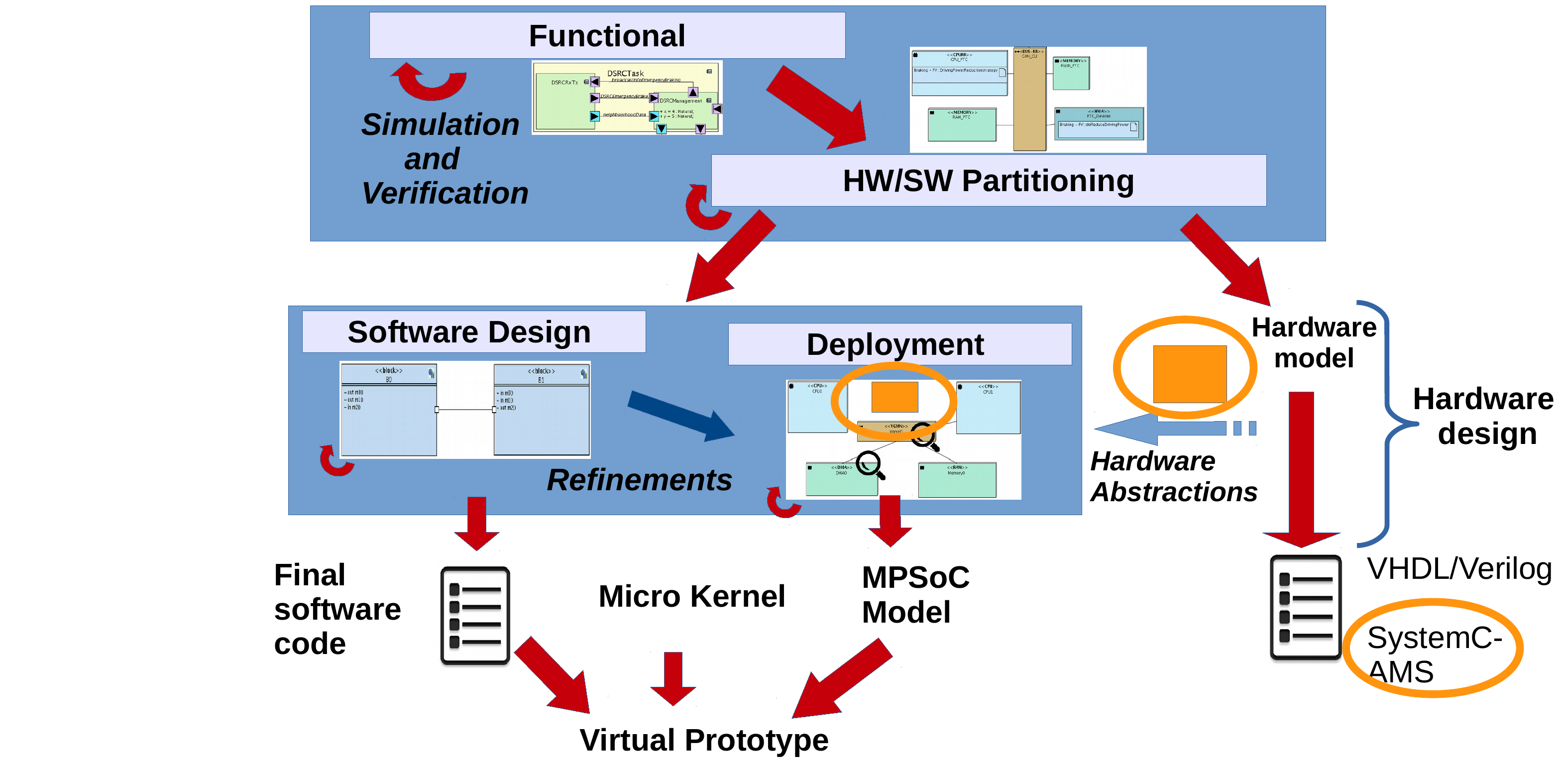}
\caption{Hardware/Software partitioning and Code generation for MPSoC platforms \cite{MODELSWARD2019}}
\label{method_w_hw}
\end{figure*}

Our modeling framework is the free and open-source software called TTool \cite{TTool}. TTool relies on SysML to propose two abstract modeling levels: 
(i) HW/SW partitioning and (ii) software design and deployment \cite{GP-NOTERE-11}.
Software tasks for the partitioning model are captured within the functional abstraction level, and software
tasks used in deployments are captured in the software design abstraction level.
In both levels, the computation part of tasks is then deployed to processors and hardware accelerators,
and the communication and storage parts are deployed to buses and memories. 
In the software deployment level, so-called \textit{deployment diagrams} 
are used to capture the allocation of software components onto a MPSoc platform. 
The tool chain which generates a cycle/bit accurate MPSoC virtual prototype built from SoCLib \cite{SOCLIB} components is explained in \cite{DGLA:ERTSS-16}. 
Figure \ref{method_w_hw}, stemming from \cite{MODELSWARD2019}, shows the overall 
design flow.
TTool was recently extended with support for modeling CPS \cite{MODELSWARD2019}, and validated for smaller examples but not yet whether the 
methodology scales to larger case studies. 

\subsection{SystemC Extensions for AMS}
\begin{figure}[h]
    \centering
    \includegraphics[width=0.47\textwidth]{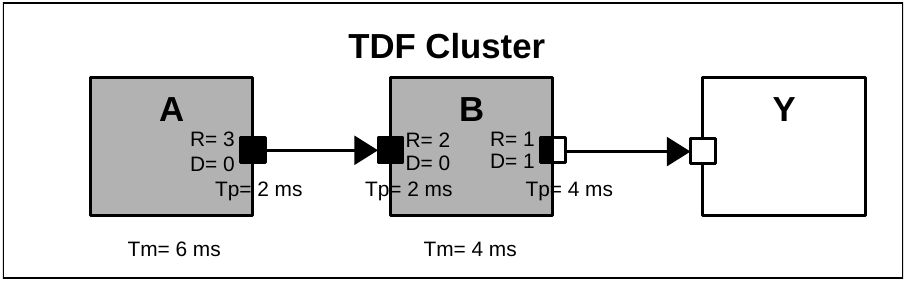}
    \caption{TDF Cluster}
    \label{fig:tdf_cluster}
\end{figure}

SystemC \cite{SYSTEMC} relies on a \textit{Discrete Event} (DE) simulation kernel. 
 \textit{Timed Data Flow} (TDF) is one of the (main) MoC of SystemC-AMS. It adds to SystemC support for signals where continuous data values are sampled with a constant time step. 
 A TDF module is described by an attribute representing a \textit{time step} and a so-called \textit{processing function}.
The time step is associated to a time period during which the processing
function should be executed. 
The processing function corresponds to a mathematical
function depending on both inputs and internal states. 
TDF modules interact with discrete parts using converter ports and have the following attributes:
\begin{enumerate}
    \item Module Timestep (\textbf{Tm}) denotes the period during which the module is activated. A module will be activated only if there are enough samples available at its input ports. 
    \item Rate (\textbf{R}). Each module will read or write a fixed number of data samples each time it is activated, annotated to the port as \emph{port rate}.
    \item Port Timestep (\textbf{Tp}) denotes the period during which each port of a module will be activated. It also denotes the time interval between two samples being read or written.
    \item Delay (\textbf{D}). A delay can be assigned to a port and will make the port handle a fixed number of samples at each activation, and read or write them in the following activation of the port. 
\end{enumerate}
Figure~\ref{fig:tdf_cluster} shows a cluster, where the DE modules are represented as white blocks, TDF modules as gray blocks, TDF normal ports as black squares, TDF converter ports as black and white squares, DE ports as white squares and TDF signals as arrows. 

\subsection{MPSoC Virtual Prototype}
In case software code is also deployed, an MPSoc platform
containing processors / buses / memories must also be generated, as well as a description of mapping of tasks and other software objects provided.
SoCLib offers a way to describe Multi-Processor System-on-Chip platforms with semantics based on the shared memory paradigm. Components can be \textbf{initiators} issuing requests (typically CPUs and hardware accelerators), or \textbf{targets} answering to requests (e.g. RAM).

In order to combine SoCLib specification with SystemC-AMS components, we have defined generic adaptor modules that can connect SystemC-AMS components to Virtual Component Interface (VCI) \cite{VCI} interfaces. 
The main idea for the integration of SystemC-AMS and SoCLib components into TTool is that analog components act as \textbf{targets} for the SoCLib initiator digital components (CPUs, DMA, \ldots). 

An adapter is modeled as a \textbf{general-purpose input/output} {(GPIO)} target component, following the modeling rules for writing cycle-bit precise SystemC simulation models for SoCLib described below. GPIO components are visible in the deployment diagram, and, like the other VCI components, their interconnection to the central VCI interconnect is represented by an arc.
The generated top cell is thus composed of SoCLib modules and interfaces to the SystemC-AMS clusters. 

\subsection{Simulation}
Due to their different Model of Computation, AMS components require to execute their simulated behavior apart from the rest of the system, but regularly synchronize with the digital platform.
The SystemC kernel is thus \textbf{controlling} the AMS kernel which
runs continuously until it is interrupted 
by access to a converter port by a TDF cluster. 
Since model-driven approaches expect to ideally provide model validation \textbf{before} code generation (and thus simulation), we propose a way to statically identify synchronization problems \cite{RodrigoMastersThesis}.

\section{Questions raised}\label{contribution}
Until now, applications modeled with the recent extension of TTool 
ran basically on mono processors, even if, as mentioned in \cite{RAPIDO2019}, 
SoCLib is designed for MPSoC virtual prototypes.
\begin{figure*}[p]
\centering
\includegraphics[width=1.0\linewidth]{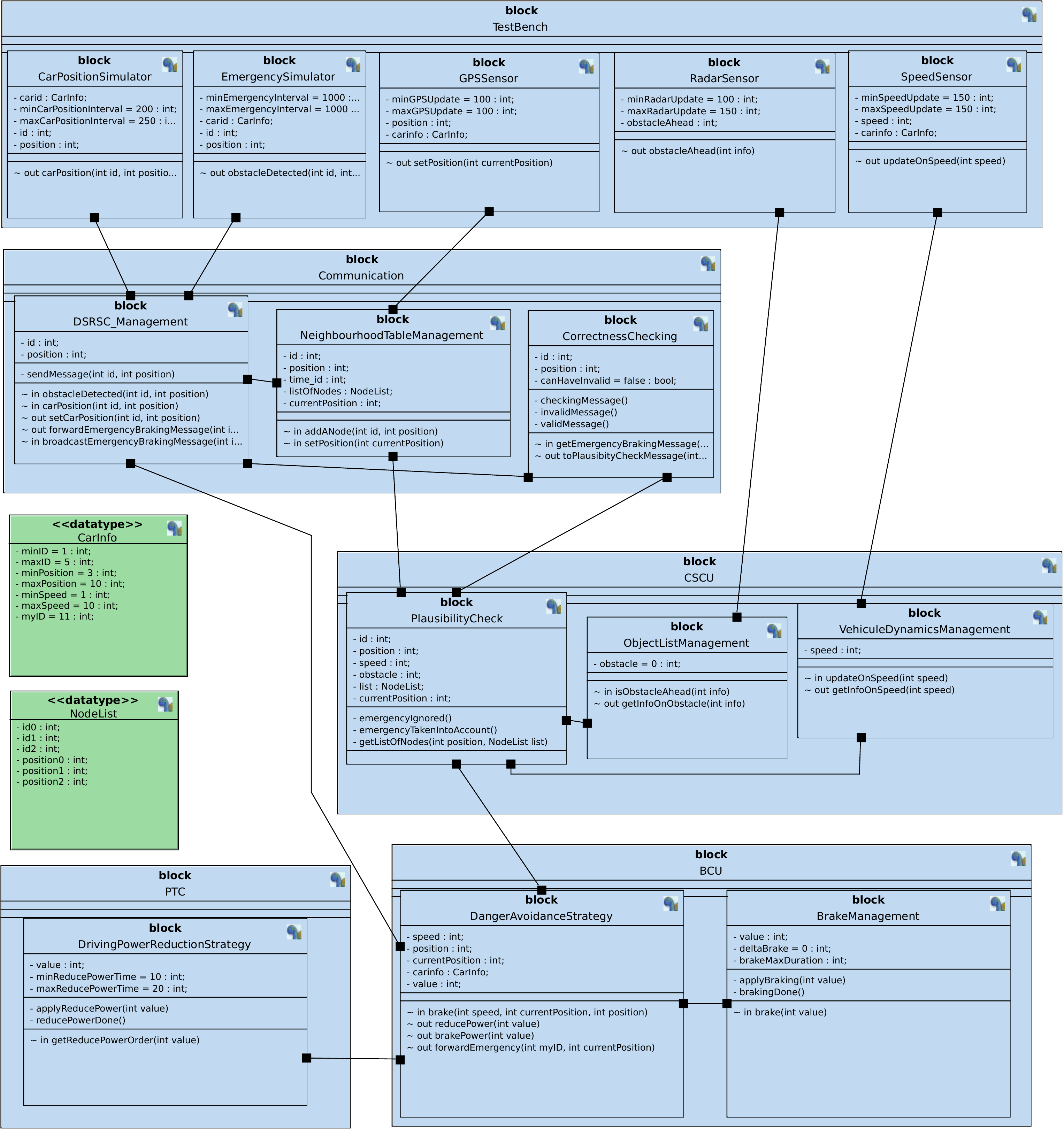}
\caption{Automotive case study: software design block diagram \cite{DASIP2016}}
\label{fig:erts_bd}
\end{figure*}
But does this also hold for multiple sensors connected by
multiple GPIO and for a MPSoC with multiple processors? 
The rover application in \cite{MODELSWARD2019} contained only two 
sensors, each came with its GPIO with one input and one output port,
and a single processor.
Larger applications feature several GPIO, more ports,
each accessing the central interconnect as a target,
generating additional traffic.

Secondly, in the presence of multiple TDF clusters, addressed by different parts of the software running on the SoCLib MPSoC, read and write operations between the digital and analog part must be handled carefully. 
Not only must causality between TDF and DE be respected
as shown in \cite{RAPIDO2019,ANDR.15a}, also the semantics of accesses by several analog blocks to the same digital block and vice versa must be preserved.
Thirdly, the GPIO interface is to date limited to transmitting only one value
at a time, of a basic type.
In SystemC-AMS, it is quite complicated to pass structured data types
as parameters of a module. While
TTool already handles them issue when generating 
pure SystemC-AMS code, GPIO interfaces do not yet take them into account.  
\section{Case Study}\label{casestudy}
These problems and current attempts at their solution are illustrated by an
automotive embedded system designed in the scope of a past European
project \cite{EVITA} and the code generation for which was presented in
\cite{DASIP2016}.

Recent on-board Intelligent Transport (IT)
architectures comprise a very heterogeneous landscape of communication
network technologies (e.g., LIN, CAN, MOST, and FlexRay) that
interconnect in-car Electronic Control Units (ECUs). 
\begin{figure}[htb]\centering
\includegraphics[width=0.65\linewidth]{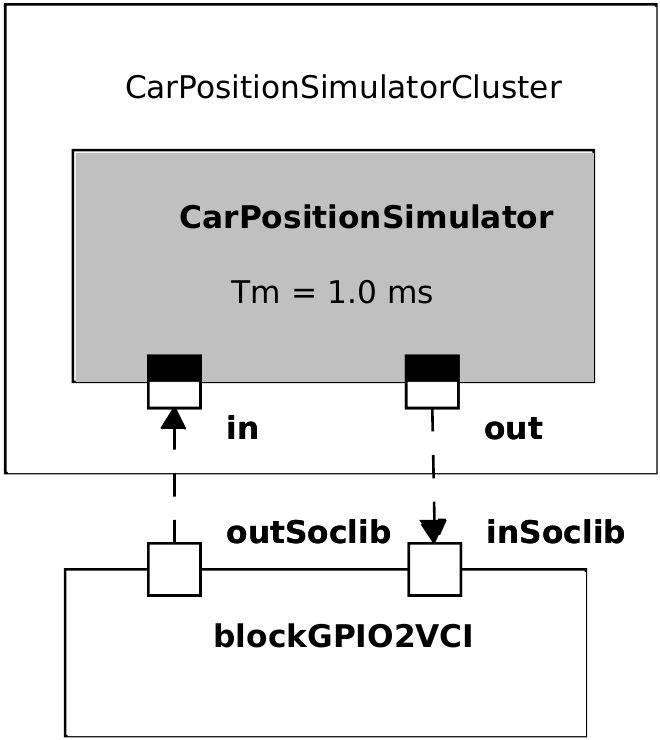}
\caption{\textit{CarPositionSimulator} sensor}\label{fig:recocyps_syscams}
\end{figure}
One of these is automatic braking \cite{EvitaD2.1}, which works essentially
as follows: an obstacle is detected by another automotive system which
broadcasts that information to neighboring cars. A car receiving
that information has to decide whether or not it is concerned. 
This decision includes a plausibility check
function that takes into account various parameters, such as the
direction and speed of the car, and also information previously received
from neighboring cars. Once the decision to brake has been taken,
the braking order is forwarded to relevant ECUs. Also, the presence of this obstacle is forwarded to
other neighboring cars in case they have not yet received this
information.

The \textit{functional view} 
comprises of a set of communicating tasks whose is described abstractly. 
Mapping then partitions the application into software and hardware.
A task mapped onto a processor will be implemented in software, and a task mapped onto a hardware accelerator is implemented in 
hardware. Tasks destined to be implemented in hardware
are either digital or analog; each task is represented on a separate panel. 
In the example, all sensors obtaining information from the environment
will be modeled as analog blocks.
\subsection{Software design}
Once the partitioning is done, the
user designs the software, for which 
functional simulation and formal verification are performed.
Figure \ref{fig:erts_bd} shows the
block diagram from \cite{DASIP2016} with
on top the five sensors, modeled as software tasks.
In the current version,
the five sensors in the \textit{TestBench} block are no longer
modeled as software tasks.
Remaining software components are grouped according to their destination ECU:
\begin{itemize}
\item \textbf{Communication ECU} manages communication with neighboring vehicles.
\item \textbf{Chassis Safety Controller ECU (CSCU)} processes emergency messages and sends orders to brake to ECUs.
\item \textbf{Braking Controller ECU (BCU)} contains two blocks:
  \textit{DangerAvoidanceStrategy} determines how to efficiently and safely reduce the vehicle speed, or brake if necessary.
\item \textbf{Power Train Controller ECU (PTC)} enforces the
  engine torque modification request.
\end{itemize}
To prototype the software components with the other elements of the destination platform (hardware components, operating system), we must map them to a model of the target system. Mapping can be performed using the deployment features introduced in \cite{DGLA:ERTSS-16}: such a \textbf{deployment diagram} is a SysML representation of hardware components, their interconnection, tasks and channels.

\subsection{Modeling sensors}
Modeling the sensors by blocks of code translated to Posix tasks
running on the MPSoC, as was practice beforehand,
oversimplified the problem.
All five sensors are thus replaced by more realistic analog models: 
five independent TDF clusters (keeping in mind that TDF still is a strong
abstraction of analog behavior).

Figure \ref{fig:recocyps_syscams} shows an AMS panel with one of the sensor clusters, the \textit{CarPositionSimulator} sensor.
In further dialogues, not shown here, parameters like rate and delay can be entered.
From these graphical information, TTool then infers, if possible, missing parameters, calculates a coherent schedule and generates SystemC-AMS code, comprising the ports, delays and interfaces \cite{RodrigoMastersThesis}.
This cluster is read by the \textit{DSRSC\_Management} block
and gives information on the car \textit{id} and \textit{position}.
Often, data structures or more than one parameter are transmitted in the
channels (here, \textit{id} and \textit{position}). Currently, they have to be transmitted one by one, basic type by basic type.
Thus, \textit{id} and \textit{position} require two sequential write operations to the out port in
the processing code and two corresponding read operations in the entry code.

We can easily model the randomized choice of an integer between 1 and 5 (\textit{id}) and between 3 and 10 (\textit{position}) stemming from the data type, on the left of Figure \ref{fig:erts_bd}.
The code of this simple processing function is shown on the right of the figure in a separate window. The \textit{write} primitive sends one integer value to the \textit{out} converter port. 

\subsection{Communication}
A library named \textit{libsyscams} 
has been provided to contain read and write primitives on the
side of the MPSoC, the \textit{read\_gpio2vci} and \textit{write\_gpio2vci} functions.
As shown above, \textit{CarPositionSimulator}
issues two random values from its output port, \textit{EmergencySimulator} does the same.

On the side of the MPSoC platform, according to TTool's semantics, the \textit{DSRSC\_Management} block non-deterministically reads from either block, or read a \textit{broadcastEmergencyBrakingMessage} from a third, the \textit{DangerAvoidanceStrategy} block. 
In the current version, the first two blocks being replaced by sensors modeled in SystemC-AMS, this semantics should be preserved. 

\begin{figure}[htb]
    \centering  
         \includegraphics[width=0.45\textwidth]{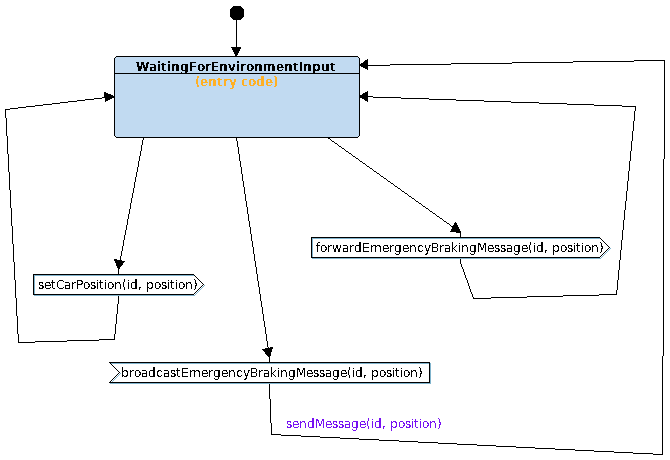}    	\caption{\textit{DSRSC\_Management} FSM containing entry code}
	\label{fig:statemachine_DSRC}
\end{figure}

In the following, we give an example of how to use \textit{libsyscams} to capture non-deterministic read operations.
Consider the finite state machine (FSM) of the \textit{DSRSC\_Management} block (Figure \ref{fig:statemachine_DSRC}).
In \cite{MODELSWARD2019}, we show how to use \textbf{entry code} that can be contained in a state to call \textit{libsyscams}.
This is the case of the \textit{WaitForEnvironmentInput} state.
Non-deterministically, either the input from \textit{CarPositionSimulator}
or \textit{EmergencySimulator} is read, whenever values are available on either.
This non-determinism, which was in the past expressed by the semantics of TTool's channels between software blocks, must now 
be reflected in the entry code of the software
block's state machine. 
If there are several parameters (here \textit{id} and \textit{position}), they must currently be read sequentially.
\begin{figure*}[htb]\centering
  \includegraphics[width=.9\linewidth]{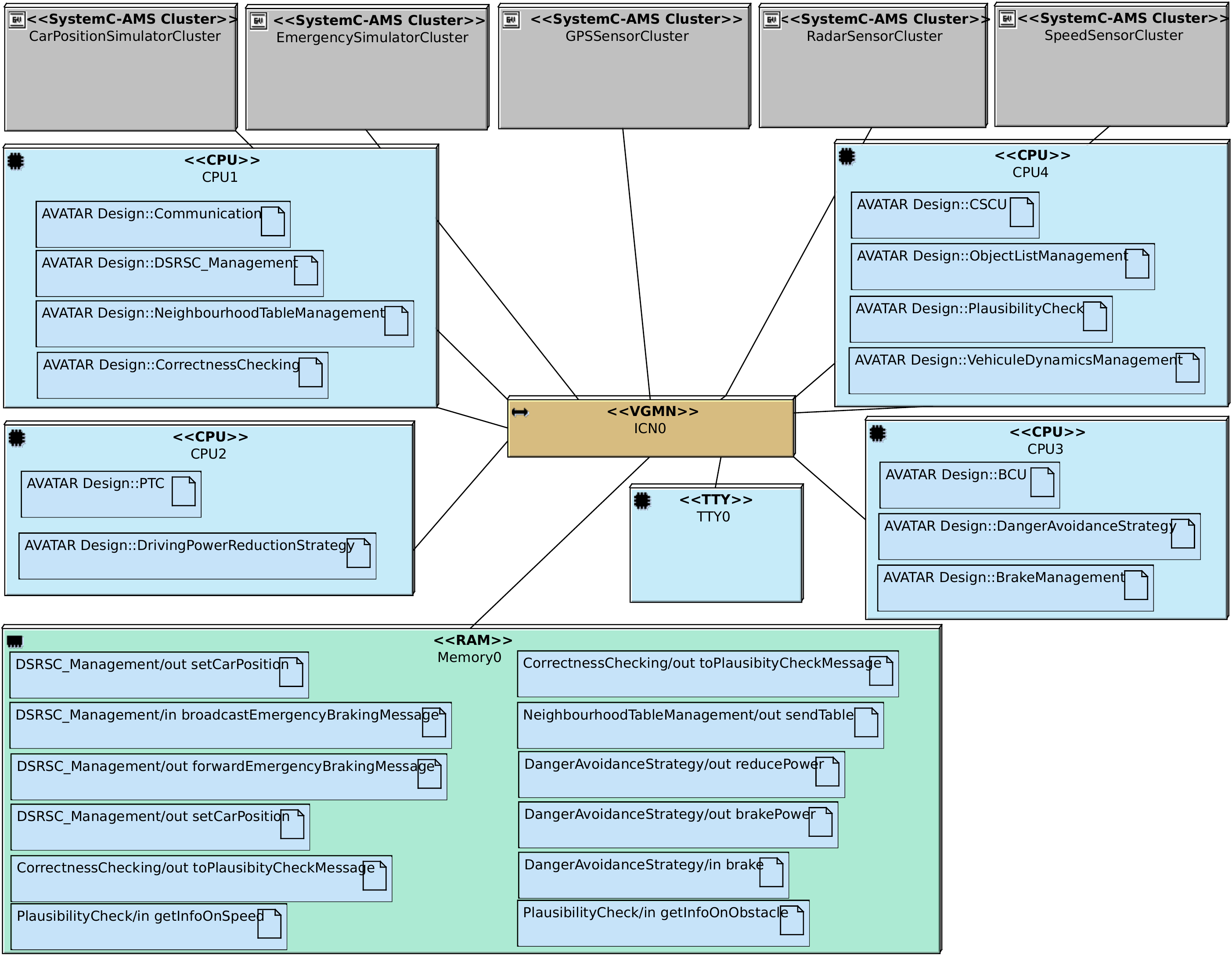}
\caption{Deployment Diagram of the Active Braking Application}\label{fig:recocyps_dd}
\end{figure*}
\subsection{Deployment}
Finally, the extended \textit{deployment diagram} (Figure \ref{fig:recocyps_dd})
 gives an overview of the mapping of software tasks and channels. The former are mapped to CPUs, the latter to on-chip memory.
For a better overview, the diagram contains sensors
as gray boxes, each corresponding to a SystemC-AMS cluster connected via a GPIO. Clicking on the box opens the corresponding SystemC-AMS panel.
A fifth CPU which contained the sensors beforehand is no longer in use.

\begin{figure}[htb!]
    \centering  
    \includegraphics[width=0.4\textwidth]{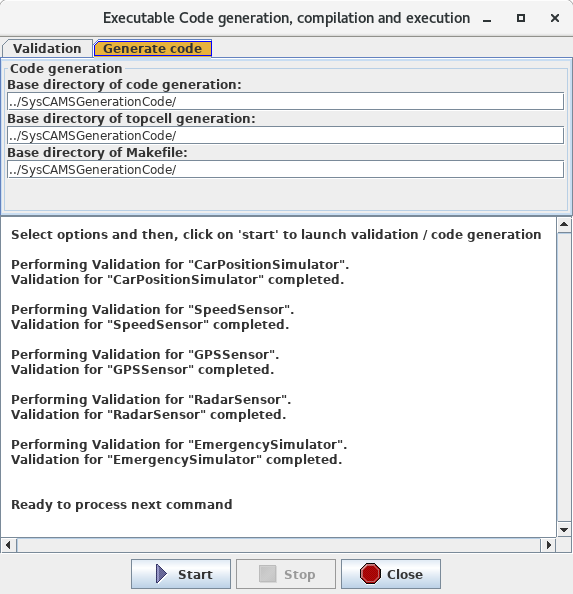}
    \caption{Validation and code generation window}
	\label{fig:validation_window}
\end{figure}

 TTool first checks the coherency of the block and port parameters before calculating a valid TDF schedule for each TDF cluster, taking into
account synchronization issues between the TDF and DE world \cite{MODELSWARD2019}. This is
done in a so-called \textit{validation} window (Figure \ref{fig:validation_window}). Once the
cluster schedule is validated, one can initiate code generation by another
mouse click.
The number of targets connected to
the central interconnect is now of 17 (five GPIO in
addition to the original 12 target modules)
and the number of initiators of 4 (one less CPU),
potentially stretching the capacity of the 
VGMN to its limits, to be explored in future experiments.
\section{Conclusion and future work}\label{conclusion}
The paper presents a case study which explores current limitations
of the AMS extensions of TTool on
a larger industrial application. 
We succeeded in running larger-scale software on a MPSoC;
exhaustive exploration and performance evaluation remains to be done.

Communication between the digital and the analog part is performed
by C entry code inserted in the state blocks, relaxing
the correct-by-construction hypothesis
of TTool. 
Non-determinism of read and write operations
should be handled more properly.
Also, it should be possible to transmit structured data types and multiple
parameters more conveniently.

The TDF models are still oversimplified; in the EVITA industrial case study however, no more detail is available. Next, we will model an application stemming from the Open Source EchOpen project \cite{ECHOPEN}, where we will have access to full implementation detail.
Finally, even if analog components tend to be unique, we plan to provide a library of parametrizable building blocks for typical components such as filters and analog/digital converters.

\bibliographystyle{ieee}
\bibliography{recocyps2019.bib}
\end{document}